\documentclass[aps,twocolumn,showpacs,showkeys]{revtex4}
\usepackage{epsfig}
\usepackage{mathrsfs}
\usepackage{epstopdf}
\usepackage{graphicx}

\newcommand{\bastar}{\begin{eqnarray*}}
\newcommand{\eastar}{\end{eqnarray*}}
\newskip\humongous \humongous=0pt plus 1000pt minus 1000pt

\newif\ifdtup

\relax
\newcommand{\be}{\begin{equation}}
\newcommand{\ee}{\end{equation}}
\newcommand{\bea}{\begin{eqnarray}}
\newcommand{\eea}{\end{eqnarray}}
\newcommand{\X}{{\vec X}}
\newcommand{\pro}{\partial}
\newcommand{\n}{\hat n}
\newcommand{\mn}{{\mu\nu}}
\newcommand{\oneg}{\displaystyle\frac{1}{g}}

\newcommand{\C}{{\vec C}}
\newcommand{\F}{\vec F}

\newcommand{\hF}{\hat F}
\newcommand{\tC}{\tilde C}

\newcommand{\A}{{\vec A}}
\newcommand{\hA}{{\hat A}}
\newcommand{\valpha}{{\vec \alpha}}

\newcommand{\hn}{{\hat n}}
\newcommand{\hD}{{\hat D}}
\newcommand{\dfrac}{\displaystyle\frac}
\newcommand{\ba}{\begin{array}}
\newcommand{\ea}{\end{array}}

\newcommand{\nn}{\nonumber}

\newcommand{\bD}{\bar D}

\begin{document}
\title{Dimensional Transmutation by Monopole Condensation in QCD}
\bigskip
\author{Y. M. Cho}
\email{ymcho@unist.ac.kr}
\affiliation{School of Electrical and Computer Engineering \\
Ulsan National Institute of Science and Technology, Ulsan 689-798 \\
and \\
School of Physics and Astronomy \\ 
Seoul National University, Seoul 151-747, Korea}

\begin{abstract}
We compare two competing conjectures of the color confinement in QCD,
the monopole condensation and the Abelian dominance, and show that it 
is the monopole condensation which is responsible for the confinement. 
To demonstrate this we present a new gauge invariant integral expression 
of the one-loop QCD effective action which has no infra-red divergence. 
With this we show that, just as the GSO-projection restores the supersymmetry 
and modular invariance in NSR string theory, the color reflection invariance 
(``the C-projection'') assures the gauge invariance and the stability of 
the monopole condensation. This establishes the monopole dominance in 
QCD. In doing so we point out critical defects in the calculation of the 
Savvidy-Nielsen-Olesen (SNO) effective action. 
\end{abstract}
\maketitle

\section{Introduction}

The confinement problem in quantum chromodynamics (QCD) is probably 
one of the most challenging problems in theoretical physics. There 
have been many conjectures on the confinement mechanism, but two of 
them are outstanding. One is the monopole condensation, the other is 
the Abelian dominance \cite{nambu,prd80,prl81,thooft,kron,suzu}. It has long been 
argued that the confinement in QCD can be triggered by the monopole 
condensation \cite{nambu,prd80,prl81}. Indeed, if one assumes the monopole 
condensation, one can easily argue that the ensuing dual Meissner 
effect could guarantee the confinement of color. To prove the monopole 
condensation, however, one must first separate the monopole potential 
from the QCD potential gauge independently. 

On the other hand the Abelian dominance proposed by 'tHooft 
asserts that the ``Abelian part'' of QCD must be responsible 
for the confinement \cite{thooft,kron,suzu}. The justification 
of this is that the colored (non-Abelian) gluons can not 
play any role in the confinement, because they (just like the 
quarks) describe the colored gluons which themselves have to be 
confined. So, only the neutral (Abelian) gluons, if at all, 
could possibly contribute to the confinement. In this sense, this 
conjecture must be true \cite{thooft,prd00}. 

The Abelian dominance has become very popular and widely been 
studied in lattice QCD \cite{kron,suzu}. As it stands, however, 
it also has serious drawbacks. First of all, we must first separate 
the Abelian part to prove this conjecture, but it does not tell us 
how to do that gauge independently. The popular way to obtain 
the Abelian part is to choose the so-called ``the maximal Abelian 
gauge'' \cite{kron,suzu}. But strictly speaking, this is a gauge 
fixing. More seriously, the maximal Abelian gauge does not tell 
exactly what constitutes the Abelian part. Obviously the Abelian 
part must contain the trivial Maxwell-type Abelian potential, but 
this U(1) potential is not supposed to produce the confinement. So 
the Abelian part must contain something else. But the maximal Abelian 
gauge does not specify what is that. This means that, even if we prove 
the Abelian dominance, we can not tell what is really responsible for 
the confinement.   

Fortunately we have a gauge independent Abelian projection defined by 
the magnetic isometry \cite{prd80,prl81}. An important feature of this 
projection is that it tells exactly what is the Abelian part. It tells 
that the Abelian potential is made of two parts, the non-topological 
Maxwell-type U(1) part and the topological non-Abelian monopole part. 
This is because in this projection the Abelian potential has the full 
non-Abelian gauge degrees of freedom, in spite of the fact that it is 
Abelian. So it naturally contains the topological degrees.

Furthermore, this projection separates the monopole potential gauge 
independently \cite{prd80,prl81}. This is very important because, unlike 
the Abelian projection based on the maximal Abelian gauge, this allows 
us to test not only the Abelian dominance but also the monopole dominance, 
and thus pinpoint exactly what is responsible for the confinement.  

Implementing the gauge independent Abelian projection on lattice, the KEK 
lattice group led by Kondo recently have demonstrated that the confinement 
in QCD is due to the monopole condensation. They have shown that the confining 
force in SU(2) QCD comes from the Abelian part of the potential, but more precisely 
the monopole part of the Abelian projection \cite{kondo1,kondo2}. This clearly 
establishes (not just the Abelian dominance but) the monopole dominance in QCD. 
This remarkable result is shown in Fig 1. 

Of course, the Abelian and/or monopole dominance have been studied in lattice 
QCD before \cite{kron,suzu,degr,born}. But these studies were based on gauge dependent 
separations of Abelian and/or monopole potentials, so that their results were 
gauge dependent. In contrast, the recent KEK calculations are explicitly gauge 
independent, because they are based on the gauge independent Abelian projection. 

\begin{figure}
\begin{center}
\psfig{figure=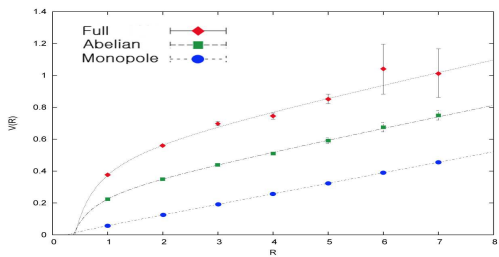, height=3.5cm, width=7cm}
\end{center}
\caption{\label{Fig. 1} The lattice QCD result which establishes the monopole 
dominance in SU(2) QCD. Here the confining potentials  marked by red, green, 
and blue points are obtained with the full SU(2) potential, the Abelian 
(restricted) potential, and the monopole potential, respectively.}
\end{figure}
  
A natural way to establish the monopole condensation in QCD is 
to show that the quantum fluctuation triggers a phase transition 
similar to the dimensional transmutation observed in massless 
scalar QED \cite{cole}. There have been many attempts to demonstrate 
this. Savvidy has first calculated the effective action of SU(2) 
QCD in the presence of an {\it ad hoc} color magnetic background, 
and has almost ``proved'' the magnetic condensation \cite{savv}. 

Unfortunately, the subsequent calculation repeated by Nielsen and 
Olesen showed that the effective action has an extra imaginary part 
which destablizes the magnetic condensation \cite{niel1,niel2}. 
This is known as the instability of the ``Savvidy-Nielsen-Olesen 
(SNO) vacuum'', which has led many people to believe that it is 
impossible to have a stable magnetic condensation in QCD \cite{ditt,yil}.

The origin of this instability can be traced to the tachyonic modes 
in the functional determinant of the gluon loop integral. But in 
physics we encounter the tachyons when we do something wrong. For 
example, in spontaneous symmetry breaking we have tachyons when we 
choose the false vacuum. Similarly, in Neveu-Schwarz-Ramond (NSR) 
string we have the tachyonic vacuum when we do not make the theory 
supersymmetric and modular invariant with the Gliozzi-Scherk-Olive 
(GSO) projection \cite{gso,witt}. The question here is how to remove 
the tachyonic modes in the gluon functional determinant, and how to 
justify that. 

The purpose of this paper is to calculate the one-loop QCD effective 
action in the presence of the monopole background, and establish the 
stable monopole condensation in SU(2) QCD theoretically. {\it Imposing 
the color reflection invariance (the C-parity) on the gluon functional 
determinant which assures the gauge invariance, we obtain a new 
infra-red finite integral expression of QCD effective action. From this 
we show that a stable monopole condensation takes place and becomes 
the true vacuum of QCD. In particular we show that, just as the GSO 
projection (the G-parity) removes the tachyonic vacuum in string theory, 
the C-parity in QCD removes the tachyons and assures the stability of 
the monopole condensation.} This establishes the dimensional transmutation 
by the monopole condensation in QCD.

It is well known that the SNO vacuum is not gauge invariant, so that 
it can not be the QCD vacuum even if it were stable \cite{niel2}.
In this paper we show another critical defect of the SNO vacuum, 
that it violates the parity. Since the parity is conserved in strong 
interaction this is a serious defect. Moreover, we show that the gauge 
invariant and parity conserving background is the monopole background, 
which is why we have chosen the monopole background in our calculation 
of the effective action.  

In doing so we point out that the SNO vacuum is not the QCD vacuum 
but a vacuum of an Abelian gauge theory coupled to a massless charged 
vector field which has no charge conjugation invariance, which is a 
sick theory that has nothing to do with QCD. 

The monopole condensation predicts two vacuum fluctuation modes of QCD, 
the $0^{++}$ and $1^{++}$ magnetic glueballs which are not made of 
the valence gluons. This is because the monopole condensation generates 
two scales, the correlation length of the monopole condensation 
and the penetration length of the color flux.   

The paper is organized as follows. In Section II we compare the 
'tHooft's Abelian decomposition and the gauge invariant Abelian 
decomposition for later purpose. In Section III we discuss the 
difference between the Savvidy background and the monopole background, 
and show that only the monopole background is gauge invariant and 
parity conserving. In Section IV we review the problems of the SNO 
effective action. In Section V we discuss the gauge invariant calculation 
of QCD effective action, and show how the color reflection invariance 
(the C-projection) excludes the tachyonic modes from the gluon 
functional determinant. In particular, we obtain a new integral 
expression of the one-loop QCD effective action which generates 
the stable monopole condensation. In Section VI we discuss a more 
general issue, the infra-red instability of the QCD effective 
action which originates from the absence of a mass scale, and show 
how the infra-red regularization by causality can restore the 
stability. In Section VII we emphasize that the perturbative 
calculation of the imaginary part can also assure the stability 
of the effective action. In Section VIII we derive the QCD effective 
potential and show that the QCD vacuum is given by the monopole 
condensation. This establishes the dimensional transmutation by
the monopole condensation in QCD. Finally in Section IX we discuss 
the physical implications of our result. 

\section{Abelian Decomposition: A review}

To establish the Abelian dominance, we must know what is the Abelian 
part of QCD which is responsible for the Abelian sub-dynamics. To do 
that we have to make the Abelian projection. There are two competing 
proposals for the Abelian projection. 

The popular Abelian projection proposed by 'tHooft is to remove as many 
non-Abelian degrees of freedom as possible by partially fixing the gauge, 
and to reduce QCD to an Abelian gauge theory coupled to the colored 
gluons \cite{thooft,kron}. To explain how this proposal is supposed to 
work, consider the SU(2) QCD for simplicity. Let $(\hn_1,\hn_2,\hn_3)$ 
be an arbitrary right-handed local orthonormal basis, and choose 
$\hn=\hn_3$ to be the Abelian (i.e., neutral) direction (which 
we can do without loss of generality). Now, let
\bea
&\A_\mu= A_\mu^1 \hn_1+ A_\mu^2 \hn_2 + A_\mu \hn
=\mathscr A_\mu+\mathscr X_\mu,  \nn\\
&{\mathscr A}_\mu=A_\mu \hn,
~~~~{\mathscr X}_\mu=A_\mu^1 \hn_1+ A_\mu^2 \hn_2,
\label{tdec}
\eea
and interprete ${\mathscr A}_\mu$ and ${\mathscr X}_\mu$ as the 
neutral and colored potentials, assuming the colored part transforms 
gauge covariantly. In this case $A_\mu$ has the residual U(1) gauge 
degrees which leaves $\hn$ invariant, which makes QCD effectively  
an Abelian gauge theory which has the colored gluons as a source. 

Here, however, we have to tell how to select the Abelian part 
${\mathscr A}_\mu$. A popular conjecture to do that is to impose the 
so-called maximal Abelian gauge condition,
\bea
&\hat {\mathcal D}_\mu {\mathscr X}_\mu=0,
~~~~\hat {\mathcal D}_\mu= \pro_\mu+g {\mathscr A}_\mu \times.
\label{mag}
\eea
The logic behind this is that $\hat {\mathcal D}_\mu {\mathscr X}_\mu$ 
becomes an isovector in color direction, so that by suppressing it with 
(\ref{mag}) one could remove as many non-Abelian degrees as possible,
and thus have the Abelian (neutral) potential ${\mathscr A}_\mu$. 
This Abelian projection has been widely adopted in lattice QCD 
to demonstrate the Abelian dominance \cite{kron,suzu}. 

This logic, however, has a serious defect. To fix ${\mathscr A}_\mu$  
from (\ref{mag}), we must know ${\mathscr X}_\mu$, but obviously we need 
${\mathscr A}_\mu$ to determine ${\mathscr X}_\mu$. This is a vicious 
circle. So it can not determine ${\mathscr A}_\mu$. In fact, (\ref{mag}) 
should (as we will see) more properly be interpreted as a condition 
on ${\mathscr X}_\mu$, not on ${\mathscr A}_\mu$. More seriously, 
(\ref{tdec}) itself has the critical defect that the decomposition of 
the neutral and colored gluons is gauge dependent. 
 
The other Abelian projection is to use the Abelian isometry to project
the Abelian part \cite{prd80,prl81}. Let $\hn$ be the neutral direction 
as before. Since this automatically determines the color directions 
$\hn_1$ and $\hn_2$, we may also call $\n$ the color direction. 
To make the Abelian projection we impose the maximal Abelian isometry
on $\A_\mu$,
\bea
D_\mu \hn=(\pro_\mu+g\A_\mu \times) \hn=0,
\label{ap}
\eea
and obtain the restricted potential $\hA_\mu$ which describes 
the Abelian sub-dynamics of QCD
\bea 
&\A_\mu \rightarrow \hA_\mu= A_\mu \n -\oneg \n \times \pro_\mu \n
=\mathscr A_\mu +\C_\mu, \nn\\
&\C_\mu= -\dfrac{1}{g} \hn \times \partial_\mu \hn. 
\label{apot}
\eea
Notice that (\ref{ap}) is gauge covariant, so that this projection 
is gauge independent. We emphasize that (\ref{ap}) restricts 
the connection precisely to the connection which leaves the Abelian 
direction $\hn$ invariant under the parallel transport.

Clearly $\hA_\mu$ is different from ${\mathscr A}_\mu$. It includes 
${\mathscr A}_\mu$, but also has $\C_\mu$ which generates a non-trivial 
field strength
\bea
&\vec H_\mn=\pro_\mu \C_\nu-\pro_\nu \C_\mu 
+g \C_\mu \times \C_\nu =H_\mn \hn, \nn\\ 
&H_{\mu\nu} = -\dfrac{1}{g} \hn \cdot(\pro_\mu \hn \times \pro_\nu \hn)
= \partial_\mu \tC_\nu-\partial_\nu \tC_\mu, \nn\\
&\tC_\mu=-\dfrac{1}{g} \hn_1 \cdot \pro_\mu \hn_2.
\label{mpot}
\eea
Notice that $\tC_\mu$ is uniquely determined up to the U(1) gauge 
transformation which leaves $\hn$ invariant. 

To understand the physical meaning of this, notice that $\C_\mu$ 
describes the non-Abelian monopole, because $\hn$ defines the mapping 
$\pi_2(S^2)$ which describes the monopole topology. Indeed, when 
$\hn=\hat r$ it describes precisely the Wu-Yang monopole \cite{wu,prl80}. 
Moreover, with
\bea
&\hn_1=\cos \phi~\hat \theta-\sin \phi~\hat \phi, 
~~\hn_2=\sin \phi~\hat \theta+\cos \phi~\hat \phi,  
\eea
the corresponding $\tC_\mu$ becomes exactly the Dirac's monopole 
potential \cite{prd80,prl80}
\bea
&\tC_\mu=-\dfrac{1}{g} \hn_1 \cdot \pro_\mu \hn_2
=-\dfrac{1}{g} (1-\cos \theta)\pro_\mu \phi.
\label{dpot}
\eea
This confirms that $\hA_\mu$ includes the non-Abelian monopole potential, 
in spite of the fact that it describes the Abelian sub-dynamics.

To clarify this dual structure,  notice that 
\bea
& \hat{F}_{\mu\nu} = \partial_\mu \hat A_\nu-\partial_\nu \hat A_\mu
+ g \hat A_\mu \times \hat A_\nu \nn\\
&= (F_{\mu\nu}+ H_{\mu\nu})\hn, \nn\\
&F_{\mu\nu} = \partial_\mu A_{\nu}-\partial_{\nu}A_\mu,
~~~H_{\mu\nu} = \partial_\mu \tC_\nu-\partial_\nu \tC_\mu. 
\eea
So the ``Abelian'' field strength of the restricted potential 
is actually made of two field strenghts, $F_\mn$ and $H_\mn$, 
each of them given by Abelian potential $A_\mu$ and $\tC_\mu$. 
In other words the Abelian part of QCD is made of two potentials,
the topological Dirac-type ``magnetic'' potential $\C_\mu$ as 
well as the non-topological Maxwell-type ``electric'' potential 
$\mathscr A_\mu$. 

With the Abelian projection we obtain the restricted chromodynamics 
(RCD) which describes the Abelian sub-dynamics of QCD \cite{prd80,prl81}
\bea
&{\cal L}_{RCD} = -\dfrac{1}{4} \hF^2_\mn =-\dfrac{1}{4} (F_\mn+H_\mn)^2.
\eea
As we will see RCD has the full SU(2) gauge freedom, in spite 
of the fact that it describes the Abelian sub-dynamics.
 
Now, we can recover the full QCD gauge potential by adding the 
colored part $\X_\mu$ to $\hA_\mu$ \cite{prd80,prl81}
\bea 
&\vec{A}_\mu = \hat A_\mu + \X_\mu 
=A_\mu \n - \oneg \n \times \pro_\mu \n+\X_\mu,     \nn\\
&\X_\mu=\dfrac1g \hn \times D_\mu \hn,~~~~\hn \cdot \X_\mu=0.
\label{adec}
\eea
This is the gauge independent Abelian decomposition of the non-Abelian 
gauge potential, also known as Cho-Duan-Ge or Cho-Faddeev-Niemi-Shabanov 
decomposition \cite{fadd,shab,zucc}.

To understand the meaning of (\ref{adec}), consider the gauge 
transformation 
\bea
\delta \A_\mu = \oneg  D_\mu \vec \alpha,
~~~~~\delta \n = - \vec \alpha \times \n,
\label{agt0}
\eea
where $\valpha$ is an infinitesimal gauge parameter. Under this 
we have
\bea
&\delta A_\mu = \oneg \n \cdot \pro_\mu \valpha,
~~~~~\delta \hat A_\mu = \oneg \hD_\mu \valpha,  \nn \\
&\delta \X_\mu = - \valpha \times \X_\mu.
\label{agt1}
\eea
This tells that $\hat A_\mu$ by itself describes an SU(2)
connection which enjoys the full SU(2) gauge degrees of
freedom. So it inherits the full topological properties of 
non-Abelian gauge theory, even though it describes only 
the Abelian sub-dynamics. This is why it contains the 
non-Abelian monopole potential, which makes RCD different 
from QED.

Furthermore $\vec X_\mu$ forms a gauge covariant vector field 
under the gauge transformation. This is because the connection 
space forms an affine space, so that any point of the connection 
space can be reached from the Abelian connection adding a gauge 
covariant vector field. Moreover it is colored ($\hn \cdot \X_\mu=0$). 
So we call $\hA_\mu$ and $\X_\mu$ the binding and valence potential 
which represents the binding and colored gluons, respectively. 

But what is really remarkable about (\ref{adec}) is that 
the decomposition is gauge independent. Once $\hn$ is chosen, 
the decomposition follows automatically, regardless of the choice 
of gauge. In particular, the separation of the monopole potential 
is gauge independent \cite{prd80,prl81}. 

So using this Abelian decomposition we can perform a truly gauge 
independent lattice calculation to test the monopole dominance. 
As we have explained the result is shown in Fig. 1, where the red, 
green, and blue dots are obtained with $\A_\mu$, $\hA_\mu$, and 
$\C_\mu$ \cite{kondo1,kondo2}. This clearly shows that it 
is the monopole which is responsible for the color confinement. 

With the decomposition (\ref{adec}), we have
\bea
&\vec{F}_{\mu\nu}=\hat F_{\mu \nu} + \hD _\mu \X_\nu -
\hD_\nu \X_\mu + g\X_\mu \times \X_\nu,
\eea
so that the Yang-Mills Lagrangian is expressed as
\bea
&{\cal L} = -\dfrac{1}{4} \F^2_{\mu \nu }=-\dfrac{1}{4}\hF_\mn^2 
-\dfrac{1}{4}(\hD_\mu\X_\nu-\hD_\nu\X_\mu)^2 \nn\\
&-\dfrac{g}{2} {\hat F}_{\mu\nu} \cdot (\X_\mu \times \X_\nu)
-\dfrac{g^2}{4} (\X_\mu \times \X_\nu)^2.
\label{ecd}
\eea
This shows that QCD can be viewed as RCD made of the binding gluons, 
which has the colored valence gluons as its source \cite{prd80,prl81}.

But we emphasize that (\ref{ecd}), strictly speaking, is different from 
QCD because it has more symmetry. This is because the decomposition 
(\ref{adec}) automatically enlarges the gauge symmetry \cite{prd01}. 
Obviously it is invariant under the gauge transformation of the active 
type, the classical (slow) gauge transformation, described by (\ref{agt0}) 
and (\ref{agt1}). But notice that it is also invariant under the following 
gauge transformation of the passive type, the quantum (fast) gauge 
transformation, 
\bea
\delta\A_\mu=\dfrac{1}{g}D_\mu \vec{\alpha},~~~~\delta \hn=0,
\label{pgt0}
\eea
under which one has
\bea
&\delta A_{\mu}=\dfrac{1}{g}\hat{n}\cdot D_{\mu}\vec{\alpha},
~~~~\delta \hA_\mu=\dfrac{1}{g} (\hn \cdot D_{\mu}\vec{\alpha})\hn, \nn\\
&\delta \X_\mu=\dfrac{1}{g}[D_{\mu}\vec{\alpha}
 -(\hn \cdot D_{\mu}\vec{\alpha})\hn].
\label{pgt1}
\eea
This is because, for a given $\A_\mu$, we can have different 
decompositions choosing different $\hn$. Equivalently, for a fixed 
$\hn$, we have different $\A_\mu$ which are gauge equivalent. 
So the decomposition inevitably enlarge the gauge symmetry. 

The extra gauge symmetry is a generic feature of the background 
field formalism in quantum field theory \cite{dewitt,pesk}. The 
decomposition of the field to classical and quantum parts does 
not tell us how to assign the symmetry of the theory to the decomposed 
parts, so that we have freedom to impose the symmetry to two parts. 
For this reason we call (\ref{ecd}) the extended chromodynamics (ECD).

The enlarged gauge symmetry plays an important role in ECD. 
According to the active gauge symmetry (\ref{agt1}), $\X_\mu$ is 
gauge covariant, so that the valence gluons (in principle) could 
acquire a mass under quantum correction. However, the passive gauge 
symmetry (\ref{pgt1}) makes this impossible. This is because we 
have to impose a gauge condition to fix the passive gauge degrees 
to quantize $\X_\mu$. A natural choice is the generalized Lorentz 
gauge \cite{prd01} 
\bea
\hD_\mu \X_\mu=0.
\label{pgcon}
\eea
Of course we could choose other gauges (e.g., the Coulomb gauge), 
but the point here is that the gauge condition keeps the valence 
gluons massless. This assures that ECD is physically equivalent 
to QCD, even though it has more symmetry. This point will play 
an important role when we calculate the effective action of QCD 
with the background field method in the following.

Now, we can compare two Abelian projections (\ref{mag}) and (\ref{ap}). 
Clearly ${\mathscr A}_\mu$ and ${\mathscr X}_\mu$ in (\ref{mag}) play 
the role of $\hA_\mu$ and $\X_\mu$ in (\ref{ap}), so that the maximal 
Abelian gauge corresponds to the Lorentz gauge (\ref{pgcon}). But notice 
that (\ref{pgcon}) is a condition to constrain $\X_\mu$, which is 
completely independent of $\hA_\mu$. If so, (\ref{mag}) should also 
be interpreted as a condition on ${\mathscr X}_\mu$, a consistency 
condition to keep the colored gluons massless. This strongly implies 
that (\ref{mag}) can not be used to determine the Abelian potential 
${\mathscr A}_\mu$. This is a critical defect of the Abelian projection 
based on the maximal Abelian gauge. 

Another disadvantage of (\ref{mag}) is that this tells nothing about 
the monopole. Obviously we need the monopole potential to test the 
monopole dominance, but (\ref{mag}) does not provide any information on
the monopole. In comparison (\ref{adec}) determines not only the Abelian 
potential $\hA_\mu$ but also the monopole potential $\C_\mu$ 
uniquely and gauge independently.    

An important advantage of (\ref{ap}) is that it can actually 
``Abelianize'' QCD gauge independently \cite{prd80,prl81}. To see this 
we replace the valence potential $\X_\mu$ by the complex vector 
field $X_\mu$,
\bea
&\X_\mu =X^1_\mu \hn_1 + X^2_\mu \hn_2,
~~X_\mu =\dfrac{1}{\sqrt{2}} (X^1_\mu+ i X^2_\mu),
\eea
and express the Lagrangian (\ref{ecd}) by 
\bea
&{\cal L}=-\dfrac{1}{4} G_{\mu\nu}^2
-\dfrac{1}{2}|\hat{D}_\mu{X}_\nu-\hat{D}_\nu{X}_\mu|^2
+ ig G_{\mu\nu} X_\mu^* X_\nu \nn\\
&-\dfrac{1}{2} g^2 \Big[(X_\mu^*X_\mu)^2-(X_\mu^*)^2 (X_\nu)^2 \Big] \nn\\
&= -\dfrac{1}{4}(G_{\mu\nu} + X_{\mu\nu})^2
-\dfrac{1}{2}|\hat{D}_\mu{X}_\nu-\hat{D}_\nu{X}_\mu|^2, \nn\\
&B_\mu = A_\mu +\tC_\mu,  \nn\\
&G_{\mu\nu} = \pro_\mu B_\nu-\pro_\nu B_\mu=F_\mn + H_\mn, \nn\\
&\hat{D}_\mu{X}_\nu = (\partial_\mu + ig B_\mu) X_\nu, \nn\\
& X_\mn = - i g ( X_\mu^* X_\nu - X_\nu^* X_\mu ). 
\label{abqcd}
\eea
Clearly this looks like an Abelian gauge theory coupled to the 
charged vector field $X_\mu$, except that this has two Abelian 
potentials. So the theory becomes a dual gauge theory whose Abelian 
gauge group is given by $U(1)_e \otimes U(1)_m$ \cite{prd80,prl81}. 
Moreover, in this Abelianization the topological structure 
of the non-Abelian gauge theory is retained in the dual structure, 
in the magnetic potential $\tC_\mu$.

Actually this Abelianized QCD is not Abelian, because (\ref{abqcd}) 
has the full non-Abelian gauge symmetry. To see this let
\bea
&\vec \alpha = \alpha_1~\hn_1 + \alpha_2~\hn_2 + \theta~\hn, 
~~~\alpha = \dfrac{1}{\sqrt 2} (\alpha_1 + i \alpha_2), \nn\\
&\C_\mu = - \dfrac {1}{g} \hn \times \partial_\mu \hn
= - C^1_\mu \hn_1 - C^2_\mu \hn_2, \nn\\
&C_\mu = \dfrac{1}{\sqrt 2} (C^1_\mu + i C^2_\mu).
\eea
Then (\ref{abqcd}) is invariant not only under the active gauge 
transformation described by
\bea 
&\delta A_\mu = \dfrac{1}{g} \partial_\mu \theta -
i (C_\mu^* \alpha - C_\mu \alpha^*),
~~~\delta \tilde C_\mu = - \delta A_\mu, \nn\\
&\delta X_\mu = 0,
\label{agt2}
\eea
but also under the following passive gauge transformation
described by
\bea 
&\delta A_\mu = \dfrac{1}{g} \partial_\mu \theta -
i (X_\mu^* \alpha - X_\mu \alpha^*), ~~~\delta \tilde C_\mu = 0, \nn\\
&\delta X_\mu = \oneg \hD_\mu \alpha - i \theta X_\mu.
\label{pgt2}
\eea
This tells that the Abelianized QCD not only retains the full 
non-Abelian gauge symmetry, but also has an enlarged (both 
active and passive) gauge symmetries.

Notice that in the conventional (Savvidy's) Abelianization based on 
the decomposition (\ref{tdec}), ${\mathscr A}_\mu$ and ${\mathscr X}_\mu$ 
transform (not to $B_\mu$ and $X_\mu$ but) to $A_\mu$ and 
$X_\mu'=(A_\mu^1+iA_\mu^2)/\sqrt 2$ . So QCD becomes an Abelian gauge 
theory made of $A_\mu$ (without $\tC_\mu$) which is coupled to $X_\mu'$. 
Adopting this view Savvidy integrated out $X_\mu'$ to obtain the QCD 
effective action, treating $A_\mu$ and $X_\mu'$ as the classical (slow) 
and quantum (fast) fields. 

This view, however, has critical defects. First of all, the separation 
of $\A_\mu$ to ${\mathscr A}_\mu$ and ${\mathscr X}_\mu$ is {\it ad hoc}, 
it is not gauge independent. Moreover ${\mathscr A}_\mu$ does not have 
the full gauge degrees of freedom, nor does ${\mathscr X}_\mu$ transform 
covariantly. So the decomposition (\ref{tdec}) has no objective meaning. 

Worse, the Abelian part here becomes exactly the Maxwell-type U(1) 
gauge theory. This trivializes QCD to QED which has the massless 
colored gluons as its source, which fails to take into account the 
dual structure of the Abelian sub-dynamics of QCD. In particular, this 
completely neglects the role of non-Abelian monopole, an essential 
feature of QCD. Clearly this is not QCD. 

\section{Savvidy Background versus Monopole Background}

An important lesson from the above discussion is that there are 
actually two sources of magnetic background in QCD, the Maxwell-type 
$F_\mn$ and the Dirac-type $H_\mn$, because the restricted potential 
has a dual structure. So we have to specify which background we must 
choose and explain why we must do so when we calculate the effective 
action. The old calculations based on (\ref{tdec}) completely missed 
this point because they had only $F_\mn$.   

To understand why this is so important, consider the gauge transformation
which inverts the color direction which we call the color reflection,
\bea
(\hn_1,\hn_2,\hn_3) \rightarrow (\hn_1,-\hn_2,-\hn).
\label{cref}
\eea
Under this we have
\bea
&\hA_\mu=\mathscr A_\mu +\C_\mu 
\rightarrow \hA_\mu^{(c)}=-\mathscr A_\mu +\C_\mu, \nn\\
&\X_\mu=X^1_\mu \hn_1 + X^2_\mu \hn_2 
\rightarrow \X_\mu^{(c)} =X^1_\mu \hn_1 -X^2_\mu \hn_2.
\label{cj1}
\eea
Notice that the electric and magnetic parts in $\hA_\mu$
transform oppositely. 

In the Abelianized QCD this translates to
\bea
&A_\mu \rightarrow A_\mu,~~~~\tC_\mu \rightarrow -\tC_\mu,  \nn\\
&X_\mu =\dfrac{1}{\sqrt{2}} (X^1_\mu+ i X^2_\mu)  
\rightarrow X_\mu^*=\dfrac{1}{\sqrt{2}} (X^1_\mu- i X^2_\mu), \nn\\
&D_\mu X_\mu=[\pro_\mu+ig(A_\mu+\tC_\mu)] X_\mu  \nn\\
&\rightarrow (D_\mu X_\mu)^*=[\pro_\mu+ig(-A_\mu+\tC_\mu)] X_\mu^*.
\eea
Clearly this tells that the red gluon $X_\mu$ changes to the blue 
gluon $X_\mu^*$ (and vise versa), and the monopole becomes 
anti-monopole, under the color reflection. As importantly the gauge 
interaction of $A_\mu$ changes the signature but that of $\tC_\mu$ 
remains invariant. Furthermore, $F_\mn$ and $H_\mn$ transform oppositely 
under the color reflection.

Obviously this color reflection which we call ``the C-parity'' is 
exactly the color charge conjugation which changes the color of 
the valence gluons. But what is important about this C-parity is 
that it must be a symmetry of QCD, because (\ref{cref}) is a gauge 
transformation. In particular, this C-parity guarantees the non-Abelian 
gauge invariance in the Abelianized QCD \cite{prd80,prl81}.

Now, since the Maxwell potential in QED has negative charge conjugation, 
$A_\mu$ must have negative C-parity. This means that $A_\mu$ and $\tC_\mu$ 
(and $F_\mn$ and $H_\mn$) in (\ref{abqcd}) have negative and positive 
C-parity, respectively. This immediately tells that the magnetic background 
made of $F_\mn$ is not gauge invariant. This, of course, was the problem 
of the SNO vacuum \cite{niel1}.
    
As importantly this assures that the monopole background made of 
$H_\mn$ is gauge invariant, because it has positive C-parity. 
So we do not have to construct ``the Copenhagen vacuum'' to obtain 
a gauge invariant background \cite{niel2}. QCD already has a natural 
candidate of gauge invariant background, the monopole background. 
And this is what we need for the dual Meissner effect.

The two potentials $A_\mu$ and $\tC_\mu$ have another important 
difference. Consider the space inversion (the parity P) 
$\vec x \rightarrow -\vec x$. Under this $A_\mu$ behaves as an 
ordinary vector, so that it has negative parity. But in spherical 
coordinates this space inversion changes $(r,\theta,\phi)$ to 
$(r,\pi-\theta,\pi+\phi)$, so that we must have 
\bea
&\hn_1=\cos \phi~\hat \theta-\sin \phi~\hat \phi
\rightarrow \hn_1^{(p)}=-\cos \phi~\hat \theta-\sin \phi~\hat \phi,  \nn\\
&\hn_2=\sin \phi~\hat \theta+\cos \phi~\hat \phi
\rightarrow \hn_2^{(p)}= -\sin \phi~\hat \theta+\cos \phi~\hat \phi, \nn\\
&\tC_\mu=-\dfrac1g \n_1\cdot \pro_\mu \n_2
=-\dfrac1g (1-\cos \theta)\pro_\mu \phi  \nn\\
&\rightarrow \tC_\mu^{(p)}=-\dfrac1g (1+\cos \theta)\pro_\mu \phi.
\label{sinv}
\eea
This shows that the monopole becomes an anti-monopole under the space 
inversion. This might be somewhat unexpected. But this is because 
$\hat r$ changes to $-\hat r$ under the space inversion, so that the 
monopole charge $\pi_2(S^2)$ changes the signature. 

To understand the physical meaning of this, notice that the monopole 
potential $\C_\mu$ in (\ref{apot}) is not sensitive to the signature 
of $\hn$. This means that in QCD there is no distinction between monopole 
and anti-monopole. In fact the monopole charge in QCD is determined 
only up to the Weyl reflection (i.e., the color reflection), so that 
monopole and anti-monopole are gauge equivalent to each other \cite{plb82}. 
So the space inversion does not change the physical content of $\tC_\mu$,
which means that it should be interpreted as an axial vector which has 
positive parity. 

{\it From this we conclude that $J^{PC}$ of the electric potential $A_\mu$ 
becomes $1^{--}$ while $J^{PC}$ of the magnetic potential $\tC_\mu$ 
becomes $1^{++}$}. This tells that the background made of the electric 
potential violates not only the gauge invariance but also the parity, 
which is a conserved quantum number in strong interaction. In contrast, 
the background made of the magnetic potential is gauge invariant and 
parity conserving.   

Before we close this section we make a few remarks. First, the monopole 
background should really be interpreted as the monopole-antimonople 
background, since monopole and anti-monopole are gauge equivalent in QCD. 
This must be contrasted with Dirac's theory of monopole, where monopole 
and anti-monopole are clearly different.

Second, the Savvidy background violates the parity (as well as the gauge 
invariance). This is another critical defect, because the Savvidy vacuum 
would violate the parity which we know is a good quantum number of strong 
interaction. As far as we understand this defect has never been pointed 
out before. 

Third, our anaysis tells that physically the space inversion is 
equivalent to the isospace inversion. There is always a residual 
U(1) gauge transformation which connects two inversions. This kind 
of space-isospace locking (e.g., the spin-isospin locking) of course 
is common in monopole and skyrmion physics \cite{gold}. 

\section{SNO Effective Action: A Review}

To understand what is wrong with the old calculations of the QCD 
effective action it is important to know how they calculated it. 
So in this section we briefly review the old calculations. To obtain 
the effective action Savvidy and others have divided the classical and 
quantum parts by (\ref{tdec}), and integrated out the quantum part 
in the presence of the Savvidy background \cite{savv,niel1,ditt,yil}
\bea
&\hF_\mn^{(b)} = \bar F_\mn \n_0,
~~~~\bar F_\mn=H \delta_{[\mu}^1 \delta_{\nu]}^2, 
\label{sb}
\eea
where $\hn_0=(0,0,1)$ is a constant Abelian direction and $H$ 
is a constant chromomagnetic field in $z$-direction. 

In this case the calculation of the functional determinant of the gluon 
loop integral amounts to finding the energy spectrum of a charged vector 
field moving around a constant magnetic field. It is well-known that this 
energy spectrum is given by \cite{tsai}
\bea
&{\cal E} = 2gH (n + \dfrac{1}{2} - q S_3) + k^2,     
\label{ev}
\eea
where $S_3$ and $k$ are the spin and momentum of the vector 
fields in the direction of the magnetic field, and $q=\pm 1$ 
is the charge (positive and negative) of the vector fields. 
This is schematically shown in Fig. 2. Notice that for both
charges the energy spectrum contains negative eigenvalues 
(tachyonic modes) which violate the causality.
 
From (\ref{ev}) we obtain the integral expression of the effective 
action 
\bea
&\Delta{\cal L} = \dfrac{}{} \lim_{\epsilon \rightarrow 0}
\dfrac{\mu^2}{16 \pi^2}\int_{0}^{\infty}
\dfrac{dt}{t^{2-\epsilon}} \dfrac{gH}{\sinh (gHt/\mu^2)} \nn\\
&\times \Big[\exp (-2gHt/\mu^2 )
+ \exp (+2gHt/\mu^2) \Big],
\label{eahx}
\eea
where $\mu^2$ is a dimensional parameter. Notice that the second term 
has a severe infra-red divergence, which we can regularize with the 
standard $\zeta$-function regularization. From the generalized 
$\zeta$-function \cite{table}
\bea
&\zeta (s,\lambda)=\dfrac{}{}\sum_{n=0}^{\infty}\dfrac{1}{(n+\lambda)^s}  
= \dfrac{1}{\Gamma(s)} \int_0^{\infty} \dfrac{x^{s-1} \exp(-\lambda x)}
{1-\exp(-x)} dx,  \nn
\label{zeta}
\eea
we have
\bea
&\Delta {\cal L} = \dfrac{}{} \lim_{\epsilon \rightarrow 0}
\dfrac{\mu^2}{8 \pi^2} gH \int_{0}^{\infty} \dfrac{dt}{t^{2-\epsilon}}  \nn\\
&\times \dfrac{\exp (-3gHt/\mu^2) + \exp (+gHt/\mu^2)}{1-\exp(-2gHt/\mu^2)} \nn\\
&= \dfrac{}{} \lim_{\epsilon \rightarrow 0}
\dfrac{(gH)^2}{4 \pi^2} (\dfrac{2gH}{\mu^2})^{-\epsilon} \Gamma(\epsilon-1) \nn\\
& \times \Big[\zeta(\epsilon-1,\dfrac{3}{2})
+ \zeta(\epsilon-1,-\dfrac{1}{2})\Big] \nn\\
&= \dfrac{11 g^2 H^2}{48 \pi^2} \big(\dfrac{1}{\epsilon} - \gamma
+1 - \ln \dfrac{2gH}{\mu^2} \big)  \nn\\
&- \dfrac{g^2 H^2}{4 \pi^2} \big(2 \zeta'(-1,\dfrac{3}{2})
- i \dfrac{\pi}{2}\big),
\label{zetareg}
\eea
where $\zeta'=\dfrac{d\zeta}{ds} (s,\lambda)$. With this we arrive at 
the SNO effective action \cite{savv,niel1,ditt}
\bea
&{\cal L}_{eff}=-\dfrac{H^2}{2} -\dfrac{11g^2 H^2}{48\pi^2}
(\ln \dfrac{gH}{\mu^2}-c) + i \dfrac {g^2 H^2} {8\pi}, \nn\\
&c=1-\ln 2 -\dfrac {24}{11} \zeta'(-1, \frac{3}{2})=0.94556... ,
\label{snoea}
\eea
which contains the well-known imaginary part which destablizes
the SNO vacuum \cite{niel1}. Obviously the imaginary part originates 
from the tachyonic eigenstates.

\begin{figure}[t]
\psfig{figure=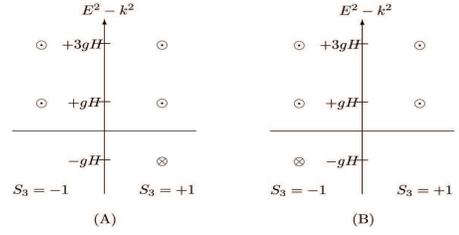, height=3cm, width=6cm}
\caption{\label{Fig. 2} The energy eigenvalues of charged vector 
fields in a constant magnetic field. (A) and (B) show the eigenvalues 
of positively and negatively charged ($q=\pm 1$) vector field. In both 
cases the eigenvalues contain tachyonic modes when $k^2<gH$.} 
\end{figure}

There have been many attempts to restore the stability to the SNO 
vacuum \cite{niel2,ditt,yil}. Actually we can remove the imaginary part 
calculating the effective action using different methods, for example, 
with the infra-red regularization by causality \cite{prd02,jhep05}. 
But the difficult part is to explain exactly what is wrong with the 
conventional (Savvidy's) calculations and the $\zeta$-function 
regularization.

Of course, this instability is not the only problem of the SNO vacuum. 
A more serious problem is that it is not gauge invariant, because it 
is not invariant under the color reflection. Furthermore, it violates 
the parity, as we have pointed out. So the SNO vacuum can not 
be identified as the QCD vacuum, even if it can be made stable.

Because of this Nielsen and Olesen have introduced ``the Copenhagen 
vacuum'', a randomly oriented blockwise Savvidy background, and 
conjectured that such background could provide a gauge invariant 
(and thus hopefully stable) QCD vacuum \cite{niel2}. Unfortunately 
the Copenhagen vacuum is not gauge invariant either, because we 
can not restore the gauge invariance by violating it randomly 
blockwise.  
                           
\section{Gauge Invariant Calculation of QCD Effective Action}

The above discussion tells that the old calculations of QCD effective 
action which renders the SNO vacuum have many defects. For example, 
the decomposition of the classical and quantum parts (\ref{tdec}) is 
not gauge independent. But this is not a serious defect, because in 
practical calculations one may always choose a convenient gauge. 

There are two critical defects in the old calculations. 
First, the Savvidy background is neither gauge invariant nor 
parity conserving, as we have emphasized. The gauge invariant 
and parity conserving background is the monopole background, 
but the old calculations based on (\ref{tdec}) could not choose 
this background because it has no place for the monopole. 

Second, the gauge invariance is completely overlooked in the 
calculation of the gluon functional determinant. In particular, 
the color reflection invariance (the C-parity) is not correctly 
implemented in the old calculations. And the instability of 
the SNO vacuum originates from this mistake.

To show how to calculate the effective action correctly, we adopt 
the gauge independent Abelian decomposition (\ref{adec}) and 
integrate out the colored gluons in the presence of the gauge 
invariant monopole background. The advantage of this, of course, 
is that the whole process of the calculation becomes explicitly 
gauge independent. 

To integrate out $\X_\mu$ we have to fix the (quantum) gauge, and 
we choose the generalized Lorentz gauge 
\bea
&\vec F=\hD_\mu \X_\mu=0,
~~~{\cal L}_{gf}=- \dfrac{1}{2\xi} (\hD_\mu \X_\mu)^2.
\label{gc}
\eea
The corresponding Faddeev-Popov determinant can be expressed by
\bea
&M_{ab}^{FP} =\dfrac {\delta F_a}{\delta \alpha^b} 
\simeq (\hD_\mu D_\mu)_{ab}-\hn_a (\hn \cdot \hD_\mu D_\mu)_b \nn\\
&=\Pi_{ac} (\hD_\mu D_\mu)_{cb},  
~~~~\Pi_{ab}=\delta_{ab}-\hn_a\hn_b.
\label{fpd}
\eea
Notice that $\Pi_{ab}$ is the projection operator which projects 
out the $\hn$ component. With this we have
\bea
&\exp ~\Big[iS_{eff}(\hA_\mu) \Big] = \dfrac{}{} \int {\cal D}
\X_\mu {\cal D} \vec{c} ~{\cal D}\vec{c}^{*}  \nn\\
&\exp \Big\{~i \dfrac{}{} \int \Big[-\dfrac {1}{4} \hF_\mn^2
-\dfrac{1}{4} (\hD_\mu \X_\nu - \hD_\nu \X_\mu)^2 \nn\\
&-\dfrac{g}{2} \hF_\mn \cdot (\X_\mu \times \X_\nu)
-\dfrac{g^2}{4}(\X_\mu \times \X_\nu)^2  \nn\\
&+\vec{c}^{~*}\hD_\mu D_\mu\vec{c}
-\dfrac{1}{2\xi} (\hD_\mu \X_\mu)^2 \Big] d^4x \Big\},
\label{ea}
\eea
where $\vec c$ and ${\vec c}^{*}$ are the ghost fields which are 
orthogonal to $\hn$. Here we need only the ghosts which are orthogonal 
to $\hn$ because they come from the gauge fixing of the valence gluons 
which are orthogonal to $\hn$.

Obviously the effective action (\ref{ea}) becomes gauge invariant 
under the transformation (\ref{agt1}), if the ghost fields transform
as
\bea
\delta \vec c = - \alpha \times \vec c,
~~~~~\delta \vec c^{*} = - \alpha \times \vec c^{*}.
\eea
So the whole process of the calculation becomes gauge independent.

To demonstrate the monopole condensation, of course, we have to
choose the gauge invariant and parity conserving monopole background,
\bea
&\hF_\mn^{(b)}= \bar H_\mn \n,
~~~~\bar H_\mn=H \delta_{[m}^1 \delta_{n]}^2.
\label{mb}
\eea
This should be compared with the Savvidy background (\ref{sb}) which 
is not gauge invariant nor parity conserving. To be general, however, 
we will let $\bar H_\mn$ arbitrary but constant in the following. 

With this the gluon and ghost loop integrals give the following 
deteminants (at one-loop level)
\bea
&{\rm Det}^{-\frac{1}{2}} K_{\mu\nu,ab} =
{\rm Det}^{-\frac{1}{2}} \Big(-g_{\mu \nu}
\bD^2_{ab}- 2g \epsilon_{abc} \bar H_\mn \n^c \Big),\nn \\
& {\rm Det} M_{ab} = {\rm Det} \Big(-\bD^2_{ab} \Big),
\label{fd}
\eea
where $\bar D_\mu$ is the covariant derivative defined by the monopole
background. From this we have
\bea
\Delta S = \dfrac{i}{2} \ln {\rm Det} K - i \ln {\rm Det} M.
\label{ea0}
\eea
So the calculation of the determinants becomes a crucial
part to obtain the effective action.
                                                                                
To calculate the functional determinants, notice that
\bea
&\ln {\rm Det} K = {\rm Tr} \ln \big(-g_{\mu\nu} \bar D^2_{ab}\big)  \nn\\
&+ {\rm Tr} \ln \Big[g_{\mu\nu} \delta_{ab}
+ 2g \bar H_{\mu\nu} \big(\dfrac{N}{\bar D^2}\big)_{ab}\Big] 
= 4 ~{\rm Tr} \ln \big(-\bar D^2_{ab}\big) \nn\\
&+ {\rm Tr} \dfrac{}{}\sum_{n=1}^{\infty} \dfrac{(-1)^{n+1}}{n} 
\Big[\big(2g \bar H_{\mu\nu}\dfrac{N}{\bar D^2}\big)^n\Big]_{ab},  \nn\\
&N_{ab} = \epsilon_{abc} n_c.
\label{gdet1}
\eea
We can simplify this to
\bea
& \ln {\rm Det} K = 4~{\rm Tr} \ln \big(-\bar D^2_{ab}\big)  \nn\\
&+ {\rm Tr} \ln \Big[\delta_{ab}  
+4a^2\big(\dfrac{N}{\bar D^2}\big)^2_{ab}\Big] 
+{\rm Tr} \ln \Big[\delta_{ab}
-4b^2\big(\dfrac{N}{\bar D^2}\big)^2_{ab}\Big] \nn\\
&=\ln {\rm Det} \Big[(-\bar D^2+2iaN)(-\bar D^2-2iaN) \Big]_{ab}  \nn\\
&+\ln {\rm Det} \Big[(-\bar D^2+2bN)(-\bar D^2-2bN)\Big]_{ab}, \nn\\
&a=\dfrac{g}{2} \sqrt{\sqrt{\bar H^4+(\bar H \tilde{\bar H})^2}+\bar H^2},  \nn\\
&b=\dfrac{g}{2} \sqrt{\sqrt{\bar H^4+(\bar H \tilde{\bar H})^2}-\bar H^2}. 
\label{gdet2}
\eea
Here $(-\bar D^2+2iaN)$ and $(-\bar D^2-2iaN)$ (also $(-\bar D^2+2bN)$ 
and $(-\bar D^2-2bN)$) correspond to two spin polarization degrees of 
valence gluons ($S_3=\pm 1$). 

Notice that $a$ and $b$ represent purely magnetic and purely electric 
background of $\bar H_\mn$, so that from now on we call them magnetic 
and electric background, respectively. This should not be confused with 
the decomposition of $\hF_\mn$ to the electric ($F_\mn$) and magnetic 
($H_\mn$) part.      
 
To proceed, we have to calculate two determinants 
\bea
&(-\bD^2 \pm 2iaN),~~~~(-\bD^2 \pm 2bN).
\label{eeq}
\eea
Now, from
\bea
(-\bD^2 \pm 2iaN) \hn = 0,~~~(-\bD^2 \pm 2bN) \hn = 0,
\eea
we may assume that the eigenfunction of the determinants
are orthogonal to $\hn$,
\bea
\vec \phi = \phi_1 \hat n_1 + \phi_2 \hat n_2.
\eea
Furthermore, with
\bea
&\bD_\mu \hat n_1 = g \bar C_\mu \hat n_2,
~~~~~\bD_\mu \hat n_2 = -g\bar C_\mu \hat n_1,  \nn\\
&\bar C_\mu=\dfrac12 \bar H_\mn x^{\nu},
\eea
we can express the eigenvalue equation of the first determinant 
by
\bea
&\left( \begin{array}{cc}
\pro_\mu^2-g^2\bar C_\mu^2,~-g(\pro_\mu \bar C_\mu 
+ 2 \bar C_\mu \pro_\mu) \pm 2ia \\
g(\pro_\mu \bar C_\mu + 2 \bar C_\mu \pro_\mu) \mp 2ia,
~\pro_\mu^2-g^2\bar C_\mu^2
\end{array} \right)
\left( \begin{array}{cc}
\phi_1 \\ \phi_2
\end{array} \right)  \nn\\
&= \lambda \left(\begin{array}{cc}
\phi_1 \\ \phi_2
\end{array} \right),
\eea
which can be diagonalized to the Abelian form
\bea
&\left( \begin{array}{cc}
-\tilde D_{+}^2 \pm 2a,~~0 \\
0,~~-\tilde D_{-}^2 \mp 2a
\end{array} \right)
\left( \begin{array}{cc}
\phi_+ \\ \phi_-
\end{array} \right)
= \lambda \left(\begin{array}{cc}
\phi_+ \\ \phi_-
\end{array} \right),  \nn\\
&\phi_{\pm} = \dfrac{\phi_1 \pm i\phi_2}{\sqrt{2}},
~~~\tilde D_{\pm}^2= (\pro_\mu \pm ig\bar C_\mu)^2.
\label{evea}
\eea
Notice that $\phi_{+}$ and $\phi_{-}$ are conjugate to each other 
under the color reflection (\ref{cref}).

Similarly, we can write the eigenvalue equation of the second 
determinant as
\bea
\left( \begin{array}{cc}
-\tilde D_{+}^2 \mp 2ib,~~0 \\
0,~~-\tilde D_{-}^2 \pm 2ib
\end{array} \right)
\left( \begin{array}{cc}
\phi_+ \\ \phi_-
\end{array} \right) 
= \lambda \left(\begin{array}{cc}
\phi_+ \\ \phi_-
\end{array} \right).
\label{eveb}
\eea
From this we have
\bea
&{\rm Det} (-\bD^2 \pm 2iaN)_{ab}  \nn\\
&= {\rm Det} (-\tilde D_{+}^2 \pm 2a)(-\tilde D_{-}^2 \mp 2a),  \nn\\
&{\rm Det} (-\bD^2 \pm 2bN)_{ab}  \nn\\  
&= {\rm Det} (-\tilde D_{+}^2 \mp 2ib)(-\tilde D_{-}^2 \pm 2ib).
\label{agdet}
\eea
Again, notice that $(-\tilde D_{+}^2 \pm 2a) \phi_+$ and 
$(-\tilde D_{+}^2 \mp 2ib) \phi_+$ are conjugate to 
$(-\tilde D_{-}^2 \mp 2a) \phi_-$ and $(-\tilde D_{-}^2 \pm 2ib) \phi_-$
under the color reflection (\ref{cref}). This means that they are gauge 
equivalent to each other, so that we must have
\bea
&{\rm Det} (-\tilde D_{+}^2 \pm 2a) = {\rm Det} (-\tilde D_{-}^2 \mp 2a), \nn\\
&{\rm Det} (-\tilde D_{+}^2 \mp 2ib) = {\rm Det} (-\tilde D_{-}^2 \pm 2ib). 
\label{ccidet}
\eea
From this we have
\bea
&\ln {\rm Det} K = 2 \ln {\rm Det} (-\tilde D^2+2a)(-\tilde D^2-2a) \nn\\
&+2 \ln {\rm Det} (-\tilde D^2-2ib)(-\tilde D^2+2ib), \nn\\
&\ln {\rm Det} M = 2 \ln {\rm Det} (-\tilde D^2), 
\label{agdetx}
\eea
where $\tilde D_\mu = \pro_\mu +ig\bar C_\mu$. Notice that the determinants 
are expressed entirely by Abelian determinants, with Abelian covariant 
derivative. 

Now, (\ref{agdetx}) reproduces the well known expression of QCD effective 
action \cite{ditt,yil}
\bea
&\Delta S = i \ln {\rm Det} (-\tilde D^2+2a)(-\tilde D^2-2a) \nn\\
&+ i \ln {\rm Det} (-\tilde D^2-2ib)(-\tilde D^2+2ib)  \nn\\
&- 2i \ln {\rm Det}(-\tilde D^2),
\label{fdabx}
\eea
and
\bea
&\Delta {\cal L} =  \dfrac{}{} \lim_{\epsilon\rightarrow0}
\dfrac{1}{16 \pi^2}  \int_{0}^{\infty} \dfrac{dt}{t^{3-\epsilon}} 
\dfrac{ab t^2}{\sinh (at/\mu^2) \sin (bt/\mu^2)} \nn\\
&\times \Big[ \exp(-2at/\mu^2)+\exp(+2at/\mu^2) \nn\\
&+\exp(+2ibt/\mu^2)+\exp(-2ibt/\mu^2)-2 \Big].
\label{eaabx}
\eea
Notice that the first four terms are the gluon loop contribution, 
but the last term comes from the ghost loop. When $a=gH$ and $b=0$, 
this becomes identical to (\ref{eahx}). And here again, the second 
and fourth terms have a severe infra-red divergence. 

\begin{figure}
\psfig{figure=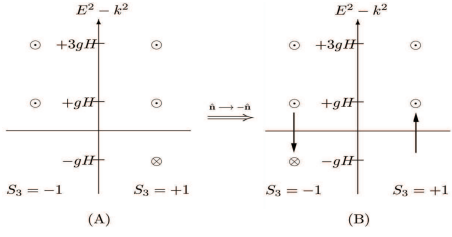, height=3cm, width=6cm}
\caption{\label{Fig. 3} The gauge invariant eigenvalues of the gluon 
functional determinant. (A) and (B) are the C-parity partners, so that 
they are gauge equivalent for each spin polarization separately. This 
shows that the negative eigenvalues in both (A) and (B) are not gauge 
invariant.}
\end{figure}
                                                                                 
We emphasize, however, this derivation of (\ref{fdabx}) has a critical 
defect. To see this remember that the gluon loop we are supposed to 
integrate out must be, not the red and blue gluon loops, but the physical 
gluon loop made of the gauge invariant combination of red and blue gluons. 
In other words the gluon loop integral ${\cal D} \X_\mu$ in (\ref{ea}) should 
really be understood as the gauge invariant integral $({\cal D} \X_\mu)_i$. 
But this important point has not been correctly taken into account in 
(\ref{fdabx}).
 
To understand this in detail, consider the case $a=gH$ and $b=0$ 
shown in Fig. 3. Clearly the color reflection (the C-parity) changes 
(A) to (B), so that they are gauge equivalent. {\it But since this 
C-parity does not change the spin of the valence gluons, we must 
require the physical eigenvalues to be invariant under the C-parity 
for each spin polarization separately,}   
\bea
&{\rm Det} (-\tilde D_{+}^2+2a) = {\rm Det} (-\tilde D_{-}^2-2a), \nn\\
&{\rm Det} (-\tilde D_{+}^2-2a) = {\rm Det} (-\tilde D_{-}^2+2a).
\label{cca}
\eea
Now, obviously the negative eigenvalues for both $S_3=+1$ in (A) and 
$S_3=-1$ in (B) do not satisfy this requirement, so that they must be 
discarded. This is the C-projection which removes the tachyonic states 
and restores the gauge invariance of the gluon loop integral. This 
neglect of gauge invariance is the critical mistake of the conventional 
calculations \cite{savv,niel1,niel2,ditt}.  

Notice that the C-parity here plays exactly the same role as the 
G-parity in string theory. It is well known that the GSO projection 
(the G-projection) restores the supersymmetry and modular invariance 
in NSR string by projecting out the tachyonic vacuum \cite{gso,witt}. 
{\it Just like the G-projection in string, the C-projection in QCD 
removes the tachyonic modes and restores the gauge invariance of the 
effective action.} 
 
Exactly the same argument applies to ${\rm Det} (-\tilde D_{+}^2 \mp 2ib)$
and ${\rm Det} (-\tilde D_{-}^2 \pm 2ib)$. Here again they are the
C-parity counterpart of each other, so that we must require
\bea
&{\rm Det} (-\tilde D_{+}^2-2ib) = {\rm Det} (-\tilde D_{-}^2+2ib), \nn\\
&{\rm Det} (-\tilde D_{+}^2+2ib) = {\rm Det} (-\tilde D_{-}^2-2ib),
\label{ccb}
\eea
for each spin polarization separately. 

This tells that (\ref{fdabx}) is incorrect. The correct functional 
determinant is given by the C-projection of it, 
\bea
& \ln {\rm Det} K= \ln {\rm Det} \Big[(-\hD^2+2iaN)(-\hD^2-2iaN) \Big]_{ab} \nn\\
&+\ln {\rm Det} \Big[(-\hD^2+2bN)(-\hD^2-2bN)\Big]_{ab} \nn\\
&= 2 \ln {\rm Det} (-\tilde D^2+2a)(-\tilde D^2+2a) \nn\\
&+2 \ln {\rm Det} (-\tilde D^2-2ib)(-\tilde D^2-2ib).
\label{agdeto}
\eea
This leads us to
\bea
&\Delta S = i \ln {\rm Det} [(-\tilde D^2+2a)(-\tilde D^2+2a)] \nn\\
&+ i \ln {\rm Det} [(-\tilde D^2-2ib)(-\tilde D^2-2ib)] \nn\\
&- 2i \ln {\rm Det}(-\tilde D^2),
\label{fdabo}
\eea
and
\bea
&\Delta {\cal L} =  \dfrac{}{} \lim_{\epsilon\rightarrow0}
\dfrac{1}{8 \pi^2}  \int_{0}^{\infty} \dfrac{dt}{t^{3-\epsilon}} 
\dfrac{abt^2/\mu^4}{\sinh (at/\mu^2) \sin (bt/\mu^2)} \nn\\
& \times \Big[\exp(-2at/\mu^2)+\exp(+2ibt/\mu^2)-1 \Big].
\label{eaabo}
\eea
This is the new integral expression of QCD effective action, 
which should be compared with (\ref{eaabx}). Obviously the 
C-projection makes (\ref{eaabo}) gauge invariant. As importantly, 
it removes the infra-red divergence of (\ref{eaabx}). 

At the first glance this might be surprising, but actually is 
not so. The gauge invariance implies the confinement which implies 
the generation of a mass parameter, which should make the theory 
infra-red finite. So it is natural that the gauge invariance 
makes (\ref{eaabo}) infra-red finite.   

When $a=0$ or $b=0$, the integration is straightforward.
For pure magnetic background we have
\bea
&\Delta{\cal L} = \dfrac{}{} \lim_{\epsilon \rightarrow 0}
\dfrac{a/\mu^2}{8 \pi^2}\int_{0}^{\infty}
\dfrac{dt}{t^{2-\epsilon}} \dfrac{\exp (-2a t/\mu^2)}{\sinh (at/\mu^2)},
\label{eaao}
\eea
so that \cite{prd02,jhep05}
\bea
&{\cal L}_{eff} = - \dfrac{a^2}{2g^2} -\dfrac{11a^2}{48\pi^2}(\ln
\dfrac{a}{\mu^2}-c).
\label{ceaa}
\eea
This is identical to the SNO effective action (\ref{snoea}),
except that it no longer has the imaginary part. This, of course,
assures the stability of the monopole condensation.

For the pure electric background we have
\bea
&\Delta {\cal L}  =  \dfrac{}{} \lim_{\epsilon \rightarrow 0}
\dfrac{b/\mu^2}{8 \pi^2}  \int_{0}^{\infty}
\dfrac{ d t}{t^{2-\epsilon}} \dfrac{\exp(+2ibt/\mu^2)}{\sin (bt/\mu^2)},
\label{eabo}
\eea
so that \cite{prd02,jhep05}
\bea
\label{ceab}
&{\cal L}_{eff} = \dfrac{b^2}{2g^2} +\dfrac{11b^2}{48\pi^2}
(\ln \dfrac{b}{\mu^2}-c) -i\dfrac{11b^2}{96\pi}.
\eea
To summarize we have
\bea
{\cal L}_{eff}=\left\{\begin{array}{ll}-\dfrac{a^2}{2g^2}
-\dfrac{11a^2}{48\pi^2}
(\ln \dfrac{a}{\mu^2}-c),~~~~~b=0 \\
~\dfrac{b^2}{2g^2} +\dfrac{11b^2}{48\pi^2}
(\ln \dfrac{b}{\mu^2}-c) \\
-i\dfrac{11b^2}{96\pi},
~~~~~~~~~~~~~~~~~~~~~~~~~~~~a=0  \end{array}\right.
\label{ceaab}
\eea
Notice that when $a=0$ the imaginary part has a negative signature, 
which implies the pair annihilation of gluons. This must be contrasted 
with the QED effective action where the electron loop integral generates 
a positive imaginary part \cite{schw,prl01}. The positive imaginary 
part in QED means the pair creation which generates the screening. 
On the other hand in QCD we must have the anti-screening to explain 
the asymptotic freedom, and the negative imaginary part is what we 
need \cite{prd02,jhep05}.  

Observe that (\ref{ceaa}) and (\ref{ceab}) are related by the 
electric-magnetic duality \cite{prd02}. We can obtain one 
from the other simply by replacing $a$ with $-ib$ and $b$ with 
$ia$. This duality, which states that the effective action should 
be invariant under the replacement
\bea
a \rightarrow - ib,~~~~~~~b \rightarrow ia,
\label{dt}
\eea
was first discovered in the QED effective action \cite{prl01}. 
But subsequently this duality has also been shown to exist in the 
QCD effective action \cite{prd02}. This tells that the duality should 
be regarded as a fundamental symmetry of the effective action of gauge 
theory, Abelian and non-Abelian. The importance of this duality is 
that it provides a very useful tool to check the self-consistency of 
the effective action. The fact that (\ref{ceaa}) and (\ref{ceab}) are 
related by the duality assures that they are self-consistent.

Clearly we can arrive at the same conclusion using the Abelianized 
QCD Lagrangian (\ref{abqcd}). The only difference is that in the 
Abelianized QCD, we integrate out the complex valence 
gluons \cite{prd02,jhep05}. 

The evaluation of the integral (\ref{eaabo}) for arbitrary $a$ and 
$b$ is not easy. Even in the ``simpler'' QED, the integration of the
effective action has been completed only recently \cite{prl01}.
But we can do the integration for arbitrary $ab\neq 0$, and prove 
that the monopole condensation is the true minimum of the effective 
potential \cite{ytmu}.     
                                                                                                                   
\section{Infra-red Regularization by Causality}

The above analysis tells that the gauge invariance (the C-projection)
resolves the instability of the SNO vacuum. But notice that this 
instability is not just the problem of the SNO vacuum but the problem 
of the QCD effective action, which originates from the infra-red 
instability of QCD itself. 

To amplify this notice that the infra-red divergence in (\ref{eahx})
was obtained with the Savvidy background, but the one in (\ref{eaabx})
was obtained with the monopole background. This tells that the 
infra-red divergence could be viewed as a generic feature of QCD 
effective action which arises in both Savvidy and monopole background,
whose origin can be traced back to the absence of a mass scale. If so, 
one may ask if there is any way to control this infra-red instability. 

Of course we have already pointed out that this is due to the mishandling 
of the gluon functional determinant. Indeed the C-projection removes this 
infra-red divergence by changing (\ref{eaabx}) to (\ref{eaabo}). But we 
may still ask if there is another way to remove the infra-red instability 
directly from (\ref{eaabx}). The answer is yes \cite{prd02,jhep05}. 

To see this remember that the effective action can be viewed as the generating 
functional of the vacuum to vacuum transition amplitude in the presence 
of the monopole background $\C_\mu$ \cite{pesk},
\bea
&\exp \big[i S_{eff} \big]
= \langle \Omega_+|\Omega_-\rangle \Big|_{\C_\mu} \nn\\
&= \dfrac{}{} \sum_{|n_i\rangle} \langle \Omega_+|n_i \rangle 
\langle n_i|\Omega_-\rangle \Big|_{\C_\mu},
\label{vtv}
\eea
where $|\Omega_{\pm}\rangle$ is the vacuum at $t=\pm\infty$, 
$|n_i\rangle$ is a complete orthonomal set of colored gluon states, 
and the gluon loop integral in (\ref{eaabx}) corresponds to the 
summation. To calculate this amplitude, of course, we must use 
physical $|\Omega_{\pm} \rangle$ and physical $|n_i\rangle$. But 
obviously the tachyonic states are not physical because they violate 
the causality, so that they should be excluded from the summation. 
The question is how. We can do that performing the integral correctly, 
with the infra-red regularization by causality \cite{prd02}.
                                                                                
To do that we start from the integral (\ref{eaabx}), and consider 
the pure magnetic background first. To impose the 
infra-red regularization by causality, we go to the Minkowski-time 
with the Wick rotation, and find \cite{prd02}
\bea
\label{eaam}
&\Delta {\cal L}=  \Delta{\cal L_+} + \Delta{\cal L_-}, \nn\\
& \Delta{\cal L_+} =  - \dfrac{}{} \lim_{\epsilon \rightarrow 0}
\dfrac{a/\mu^2}{16 \pi^2}\int_{0}^{\infty}
\dfrac{dt}{t^{2-\epsilon}} 
\dfrac{\exp (-2iat/\mu^2)}{\sin (at/\mu^2)}, \nn\\
& \Delta{\cal L_-}=-\dfrac{}{} \lim_{\epsilon \rightarrow 0}
\dfrac{a/\mu^2}{16 \pi^2}\int_{0}^{\infty}
\dfrac{dt}{t^{2-\epsilon}} 
\dfrac{\exp (+2iat/\mu^2)}{\sin (at/\mu^2)}.
\eea
In this form the infra-red divergence has disappeared, but now 
we face an ambiguity in choosing the correct contours of the 
integrals. 

Fortunately we can resolve this ambiguity imposing the causality. 
To see this notice that the two integrals $\Delta{\cal L_+}$ and 
$\Delta{\cal L_-}$ originate from two determinants $(-\tilde D^2+2a)$ 
and $(-\tilde D^2-2a)$, but the standard causality argument (with 
the familiar Feynman prescription $p^2 \rightarrow p^2-i\epsilon$) 
requires us to identify $2a$ in the first determinant as $2a-i\epsilon$ 
but in the second determinant as $2a+i\epsilon$. This tells that 
the poles in the first integral in (\ref{eaam}) should lie above the 
real axis, while the poles in the second integral should lie below 
the real axis. From this we conclude that the two integrals become 
complex conjugate to each other. This guarantees that $\Delta{\cal L}$ 
is explicitly real, without any imaginary part. Moreover the real part 
of $\Delta{\cal L_+}$ and $\Delta{\cal L_-}$ must be the same. So with 
this infra-red regularization by causality we obtain \cite{prd02}
\bea
&{\cal L}_{eff} = - \dfrac{a^2}{2g^2} -\dfrac{11a^2}{48\pi^2}(\ln
\dfrac{a}{\mu^2}-c), \nn
\eea
which is identical to (\ref{ceaa}).

For the pure electric background the infra-red 
regularization by causality, with a similar reasoning,
yields \cite{prd02,jhep05}
\bea 
&{\cal L}_{eff} = \dfrac{b^2}{2g^2} +\dfrac{11b^2}{48\pi^2}
(\ln \dfrac{b}{\mu^2}-c) -i\dfrac{11b^2}{96\pi}, \nn
\eea
which is identical to (\ref{ceab}). This confirms that we can indeed 
control the infra-red instability of the effective action by causality.
It is really remarkable that two completely independent principles, 
the gauge invariance and the causality, produce exactly the same result.

Of course, this infra-red regularization by causality could make the 
SNO vaccum stable. But this is not the point. As we have emphasized, 
the SNO vacuum can not be the QCD vacuum independent of whether it is 
stable or not. It violates the gauge invariance and the parity. Nor 
does it tell that the $\zeta$-function regularization is wrong. It 
works well with (\ref{eaabo}). This merely tells that there is a physical 
regularization which can correct the infra-red instability of (\ref{eaabx}), 
independent of what is the QCD vacuum. 

\section{Perturbative Confirmation}

There is another powerful way to calculate the imaginary part in 
the effective action. This is because in massless gauge theories 
(in particular in QCD) the imaginary part of the one-loop effective 
action is of the order of $g^2$, so that we can actually calculate 
it perturbatively \cite{sch,prd02}. Considering that the effective 
action is essentially non-perturbative, this is unexpected. But this 
has been demonstrated in both massless QED and QCD \cite{prd02,jhep05}.

In QED Schwinger has obtained the following effective action
perturbatively to the order $e^2$ \cite{schw}
\bea
&\Delta S_{QED}=\dfrac{e^2}{16 \pi^2} \int d^4p
F_{\mu\nu}(p)F_{\mu\nu}(-p) \nn\\
&\times \dfrac{}{}\int_{0}^{1} dv \dfrac{v^2 (1- v^2/3)}{(1- v^2)
+ 4m^2/p^2},
\label{qedea}
\eea
where $m$ is the electron mass. From this he observed that
when $-p^2>4m^2$ the integrand develops a pole at
$v^2=1+4m^2/p^2$ which generates an imaginary part, and
explained how to calculate it. But notice that in the massless 
limit, the pole moves to $v=1$. In this case the pole contribution
to the imaginary part is reduced by a half, and we obtain
\bea
Im ~{\cal L}_{QED} {\Big |}_{m=0} = \left\{{~0
~~~~~~~~~~~~~~ b=0,
\atop \dfrac{b^2}{48 \pi} ~~~~~~~~~~~a=0.}\right.
\eea
This is exactly what we obtain from the non-perturbative QED effective 
action in the massless limit \cite{prd02,jhep05}.

In QCD we can find the imaginary part either by calculating the 
one-loop Feynman diagrams directly, or by evaluating the integral 
(\ref{eaabx}) to the order $g^2$ using Schwinger's method. Here 
we do this with the Feynman diagrams. 

For an arbitrary monopole background there are four Feynman diagrams 
that contribute to the order $g^2$ which are shown in Fig. 4. The sum 
of these diagrams (in the Feynman gauge with dimensional regularization) 
gives us \cite{pesk}
\bea 
&\Delta S = -\dfrac{11g^2}{96 \pi^2} \int d^4p
H_{\mu\nu}(p)H_{\mu\nu}(-p) \nn\\
&\times \left[\mbox{ln}
\left(\dfrac{p^2}{\mu^2}\right) + C_1 \right],
\label{peafeyn}
\eea
where $C_1$ is a regularization-dependent constant. Clearly 
the imaginary part could only arise from the logarithmic term 
$\mbox{ln} (p^2/\mu^2)$, so that for a space-like $p^2$ (with 
$\mu^2>0$) the effective action has no imaginary part. But 
a space-like $p^2$ corresponds to a magnetic background. 
This means that (\ref{eaabx}) has no imaginary part when $b=0$, 
at least at the order $g^2$. 

\begin{figure}
\psfig{file=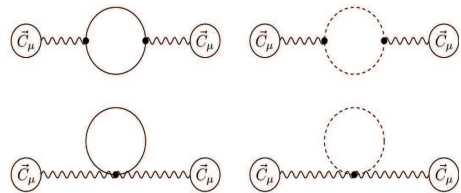, height=2.5cm, width=6cm}
\caption{\label{Fig. 4} The Feynman diagrams that contribute to the
effective action at $g^2$ order. Here the straight line and the dotted 
line represent the valence gluon and the ghost, respectively.}
\end{figure}
                       
To evaluate the imaginary part for an electric background we have 
to make the analytic continuation of (\ref{peafeyn}) to a time-like 
$p^2$. In this case the causality (again with the Feynman prescription 
$p^2 \rightarrow p^2-i \epsilon$) dictates us to have
\bea
&\mbox{ln} \Big(\dfrac{p^2}{\mu^2}\Big) \rightarrow
\lim_{\epsilon \to0} \mbox{ln} \Big(\dfrac{p^2-i\epsilon}{\mu^2} \Big) \nn\\
&=\mbox{ln} \Big(\dfrac{|p^2|}{\mu^2}\Big)- i \dfrac{\pi}{2}
~~~~~(p^2<0),
\label{causal}
\eea
so that we obtain
\bea 
Im ~\Delta {\cal L} = \left\{{~~~~0~~~~~~~~~~~~(b=0),
\atop -\dfrac{11 b^2}{96 \pi} ~~~~~~~~~(a=0).}\right.
\label{imea}
\eea
Obviously this is identical to (\ref{ceaab}). And we can reproduce 
the same result using the Schwinger's method \cite{jhep05}.
                             
It might look surprising that both the infra-red regularization 
by causality and the perturbative calculation produce the same imaginary 
part. But this is natural, because both are based on the causality.

This confirms that we can control the infra-red instability of QCD 
effective action if we treat it correctly. In particular this tells 
that the tachyonic modes in (\ref{eaabx}) are indeed unphysical mirage 
which should not have been there in the first place. They come into 
the calculation of the functional determinant by default. 

\section{Dimensional transmutation}

The effective action (\ref{ceaa}) generates the much desired
dimensional transmutation in QCD, the phenomenon Coleman and 
Weinberg first observed in massless scalar QED \cite{cole,pesk}.
To demonstrate this notice that the effective action (\ref{ceaa}) 
provides the following effective potential
\bea
V=\dfrac{a^2}{2g^2}
\Big[1+\dfrac{11 g^2}{24 \pi^2}(\ln\dfrac{a}{\mu^2}-c)\Big].
\eea
So, defining the running coupling $\bar g$ by \cite{prd02,jhep05}
\bea
\frac{\partial^2V}{\partial a^2}\Big|_{a=\bar \mu^2}
=\frac{1}{ \bar g^2},
\eea
we have
\bea
&\dfrac{1}{\bar g^2} =\frac{1}{g^2}
+\frac{11}{24 \pi^2}( \ln\frac{{\bar\mu}^2}{\mu^2}
- c + \dfrac{3}{2}),  \nn\\
&\beta(\bar\mu)= \bar\mu \dfrac{\partial \bar g}{\partial \bar\mu}
= -\frac{11 \bar g^3}{24\pi^2}.
\label{beta}
\eea
This is exactly the same $\beta$-function in perturbative SU(2) QCD 
which assures the asymptotic freedom \cite{wil}. 
                                                                                                      
In terms of the running coupling the renormalized potential is 
given by
\bea
V_{\rm ren}=\dfrac{a^2}{2\bar g^2}
\Big[1+\dfrac{11 \bar g^2}{24 \pi^2 }
(\ln\dfrac{a}{\bar\mu^2}-\dfrac{3}{2})\Big],
\eea
which generates a non-trivial local minimum at
\bea
\langle a \rangle=\bar \mu^2 \exp\Big(-\frac{24\pi^2}{11\bar g^2}+ 1\Big).
\eea
Notice that with ${\bar \alpha}_s = 1$ we have
\bea
\dfrac{\langle a \rangle}{{\bar \mu}^2} = 0.48988... 
\eea
This is nothing but the dimensional transmutation by the monopole 
condensation. The corresponding effective potential is plotted in Fig. 5,
where we have assumed $\bar \alpha_s=1$ and $\bar \mu^2=1$.
                                                                                
Nelsen and Olesen have suggested that the existence of the unstable 
tachyonic modes are closely related with the asymptotic freedom in 
QCD \cite{niel1}.  Our analysis tells that this is not true. Obviously 
the asymptotic freedom shown in (\ref{beta}) follows from the stable 
monopole condensation in the absence of tachyons.

\begin{figure}
\psfig{file=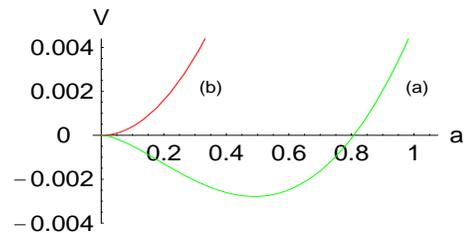, height=3cm, width=6cm}
\caption{\label{Fig. 5} The effective potential of SU(2) QCD in
the monopole background. Here (a) is the effective potential
and (b) is the classical potential.} 
\end{figure}                 
                                                               
\section{Discussion}

Establishing the dimensional transmutation by the monopole 
condensation in QCD has been extremely difficult to attain. 
Savvidy and many others had the correct idea, but made serious 
mistakes in implementing the idea. Their calculations (in 
particular their Abelian decomposition) were not gauge independent. 
This in itself may not be a serious problem, but the old calculations 
have two critical defects. First, they have chosen the wrong 
background which is neither gauge invariant nor parity conserving. 
This is the problem of the SNO background. Second, they have failed 
to impose the gauge invariance in the gluon loop integral. This is 
the problem of the infra-red instability of the effective action. 
 
The first problem arises because the old calculations neglected 
the fact that there are actually two possible magnetic backgrounds 
in QCD, the SNO background and the monopole background, but only 
the monopole background qualifies to be the QCD vacuum because 
only this background is gauge invariant and parity conserving. 
This is because they did not take into account the fact that the 
restricted potential has a dual structure. 

The second problem, the infra-red instability of the effective action, 
originates from the fact that the old calculations failed to implement 
the gauge invariance in the calculation of the gluon functional 
determinant properly. In this paper we have shown how the C-projection 
makes the functional determinant gauge invariant and removes the 
tachyonic modes which cause the effective action unstable. This 
assures the stable monopole condensation. 

The Abelian decomposition (\ref{adec}) provides a perfect setting 
for us to calculate the one-loop effective action of QCD. This 
decomposition separates not only colored potential but also the monopole 
gauge independently, and allows us to choose the gauge invariant monopole 
background. Moreover it also tells exactly how each term transforms 
under the gauge transformation (the color reflection), and tells how 
to obtain the gauge invariant effective action with the color reflection. 
So using this decomposition, we can calculate the correct effective action 
and establish the stable monopole condensation. 

{\it An unavoidable consequence of the monopole condensation is 
the existence of two magnetic glueballs, the $0^{++}$ and $1^{++}$ 
modes of the vacuum fluctuation, which are not made of the valence 
gluons.} This is because the monopole condensation induces two scales, 
the correlation length of the monopoles and the penetration length 
of the color flux \cite{prl81}. Obviously they are similar to the 
Higgs scalar and the massive photon that we have in spontaneous 
symmetry breaking, and it is very difficult to avoid these modes 
if we believe in the monopole condensation. The challenge now is 
to confirm (or disprove) the existence of these magnetic glueballs 
experimentally. 

The existence of the tachyonic modes which caused the infra-red instability 
of the effective action has been a headache, the Gordian knot, in 
QCD. It was there, and nobody knew how to resolve this puzzle. In fact
it has been thought that this infra-red instability is an essential 
characteristic, a sacred feature, of QCD which comes from the absence 
of a mass parameter. But in this paper we have shown that the tachyons 
are an unphysical mirage which should not have been there in the first 
place. They disappear when we calculate the effective action correctly. 
The gauge invariance, the infra-red regularization by causality, and 
the perturbative calculation of the imaginary part all confirm that 
there is no infra-red instability in the QCD effective action.  

As we have emphasized, in physics tachyons appear when we do 
something improper or choose something unphysical. In spontaneous 
symmetry breaking tachyons appear when we choose the false vacuum. 
Similarly, in NSR string we have the tachyonic vacuum when we do 
not impose the modular invariance and supersymmetry with the 
G-projection \cite{gso,witt}. Here we have the same situation. 
The tachyons appear when we we do not implement the gauge invariance 
to the gluon functional determinant correctly. This tells that 
there is nothing mysterious about the tachyons in QCD.

It is not surprising that the gauge invariance plays the key role
to generate the stable monopole condensation. From the beginning 
the gauge invariance has been the main motivation for the confinement. 
It is this gauge invariance which forbids any colored object from 
the physical spectrum of QCD. So it is only natural that the gauge 
invariance assures the stability of the monopole condensation, and 
thus the confinement of color. 

The Abelian dominance conjecture has been a very popular confinement 
mechanism \cite{thooft,kron}. But actually it is a misleading and 
confusing conjecture. Obviously it is hollow unless it tells what 
is the Abelian part, and the maximal Abelian gauge (\ref{mag}) 
does not tell what is that. The gauge independent Abelian projection 
(\ref{ap}) does tell what is the Abelian part, but here the Abelian 
part has a dual structure. So we have to tell which is important in 
the Abelian dominance.    

Our analysis confirms that it is the monopole which provides the 
confinement. And this has been endorsed by the recent lattice 
calculation \cite{kondo1,kondo2}. This is made possible because 
we have the Abelian projection (\ref{ap}) which projects out (not 
only the Abelian potential but also) the monopole potential gauge 
independently. 

It must be emphasized, however, that the monopole condensation should 
really be understood as the monopole-antimonopole condensation. 
This is because in QCD the monopole and anti-monopole are gauge 
equivalent, because they are the C-parity partners \cite{plb82}.

One may wonder if there is any classical monopole-antimonopole 
configuration which can suggest the stable monopole condensation.
Indeed there is. It has been shown that a pair of monopole-antimonopole 
strings becomes stable under the quantum fluctuation, if the distance 
between two strings becomes less than a critical distance \cite{plb06}. 
This strongly implies that monopole-antimonopole pairs can be stable 
against the quantum fluctuation.  
      
At this point one may ask what is wrong with the SNO effective action.
Obviously it is not the effective action of QCD. But we can certainly 
say that it is the effective action of a U(1) gauge theory coupled to 
a massless charged vector field which has no charge conjugation (color 
reflection) invariance. This is not QCD, but actually a sick theory. 
It is well known that such theory (a theory of massless charged vector 
field which has no non-Abelian gauge invariance) is ill-defined. So there is nothing 
wrong with the SNO effective action, except that it is the effective 
action of a sick theory. And the problems of the SNO vacuum are the 
symptoms of this sickness.

It is truly remarkable (and surprising) that the principles of
quantum field theory allow us to demonstrate confinement within 
the framework of QCD. There has been a proof of confinement in 
a supersymmetric QCD \cite {seib}. Our analysis shows that we can 
actually establish the existence of the confinement phase within 
the conventional QCD, with the existing principles of quantum field 
theory. This should be interpreted as a most spectacular triumph of 
quantum field theory itself.

We conclude with the following remarks: \\
1. To show that the monopole condensation is the true vacuum of 
QCD, we have to integrate the effective action (\ref{eaabo}) for 
an arbitrary chromo-electromagnetic background. This is not easy 
but doable, and we can show that indeed the monopole background 
remains a true minimum of the effective potential, at least at 
one-loop level \cite{ytmu,cho1}. \\ 
2. In this paper we have neglected the quarks. We simply remark that 
the quarks, just as in the asymptotic freedom, tend to destabilize 
the monopole condensation. But if the number of quarks are small 
enough, the condensation remains stable. In fact the stability puts 
exactly the same constraint on the number of quarks as the asymptotic 
freedom \cite{ytmu,cho1}.  \\
3. In real QCD, of course, we have to deal with SU(3). We can do this, 
and the generic feature of the SU(2) QCD, in particular the dimensional 
transmutation by the monopole condensation, remains the same \cite{cho1}. 

The calculation of the QCD effective action for $ab\neq 0$  which 
includes the contribution of the quark loop will be published 
separately \cite{cho1}.

{\bf Acknowledgements}

~~~This work is supported in part by the Basic Science Research Program 
through the National Research Foundation of Korea funded by the Ministry 
of Education, Science, and Technology (2012-002-134).


\begin{thebibliography}{99}
\bibitem{nambu} Y. Nambu, Phys. Rev. {\bf D10}, 4262 (1974); 
S. Mandelstam, Phys. Rep. {\bf 23C}, 245 (1976);
A. Polyakov, Nucl. Phys. {\bf B120}, 429 (1977).
\bibitem{prd80} Y. M. Cho, Phys. Rev. {\bf D21}, 1080 (1980).
See also Y. S. Duan and M. L. Ge, Sci. Sinica {\bf 11},1072 (1979).
\bibitem{prl81} Y. M. Cho, Phys. Rev. Lett. {\bf 46}, 302 (1981); 
Phys. Rev. {\bf D23}, 2415 (1981); W. S. Bae, Y. M. Cho, and S. W. Kimm, 
Phys. Rev. {\bf D65}, 025005 (2002).
\bibitem{thooft} G. 't Hooft, Nucl. Phys. {\bf B190}, 455 (1981). 
\bibitem{kron} A Kronfeld, G. Schierholz, and U. Wiese, Nucl. Phys. {\bf B293}, 461 (1987);
A Kronfeld, M. Laursen, G. Schierholz, and U. Wiese, Phys. Lett. {\bf B198}, 516 (1987).
\bibitem{suzu} T. Suzuki and I. Yotsuyanagi, Phys. Rev. {\bf D42}, 4257 (1990);
G. Bali, V. Bornyakov, M. Mueller-Preussker, and K. Schilling, Phys. Rev. {\bf D54}, 2863 (1996).
\bibitem{prd00} Y. M. Cho, Phys. Rev. {\bf D62}, 074009 (2000).
\bibitem{kondo1} S. Kato, K. Kondo, T. Murakami, A. Shibata, T. Shinohara, and S. Ito, 
Phys. Lett. {\bf B632}, 326 (2006).  
\bibitem{kondo2} S. Ito, S. Kato, K. Kondo, T. Murakami, A. Shibata, and T. Shinohara, 
Phys. Lett. {\bf B645}, 67 (2007); {\bf B653}, 101 (2007); {\bf B669}, 107 (2008).
\bibitem{degr} T. DeGrand and D. Toussaint, Phys. Rev. {\bf D22}, 2478 (1980). 
\bibitem{born} V. Bornyakov, H. Ichie, Y. Mori, D. Pleiter, M. Polikarpov, 
G. Schierholz, T. Streuer, H. Stuben, and T. Suzuki, Phys. Rev. {\bf D70}, 
054506 (2004); V. Bornyakov, H. Ichie, Y. Koma, Y. Mori, Y. Nakamura, 
D. Pleiter, M. Polikarpov, G. Schierholz, T. Streuer, H. Stuben, 
and T. Suzuki, Phys. Rev. {\bf D70}, 074511 (2004).
\bibitem{cole} S. Coleman and E. Weinberg, Phys. Rev. {\bf D7},
1888 (1973).
\bibitem{savv} G. K. Savvidy, Phys. Lett. {\bf B71}, 133 (1977).
\bibitem{niel1} N. Nielsen and P. Olesen, Nucl. Phys. {\bf B144}, 
485 (1978).
\bibitem{niel2} N. Nielsen and P. Olesen, Nucl. Phys. {\bf B160}, 
380 (1979); C. Rajiadakos, Phys. Lett. {\bf B100}, 471 (1981).
\bibitem{ditt} W. Dittrich and M. Reuter, Phys. Lett. {\bf B128}, 321, (1983);
{\bf B144}, 99 (1984); C. Flory, Phys. Rev. {\bf D28}, 1425 (1983);
S. K. Blau, M. Visser, and A. Wipf, Int. J. Mod. Phys.
{\bf A6}, 5409 (1991); M. Reuter, M. G. Schmidt, and C. Schubert,
Ann. Phys. {\bf 259}, 313 (1997).
\bibitem{yil} A. Yildiz and P. Cox, Phys. Rev. {\bf D21}, 1095
(1980); M. Claudson, A. Yilditz, and P. Cox, Phys. Rev. {\bf D22}, 2022
(1980); J. Ambjorn and R. Hughes, Phys. Lett. {\bf B113}, 305 (1982).
\bibitem{gso} F. Gliozzi, J. Scherk, and D. Olive, Nucl. Phys. {\bf B122}, 253 (1983).
\bibitem{witt} See, e.g., M. Green, J. Schwarz, and E. Witten, 
{\it Superstring Theory} Vol. I, Cambridge University Press, 1987; 
M. Kaku, {\it Introduction to Superstrings}, Springer-Verlag
(New York), 1988. 
\bibitem{wu}T. T. Wu and C. N. Yang, Phys. Rev. {\bf D12}, 3845 (1975).
\bibitem{prl80} Y. M. Cho, Phys. Rev. Lett. {\bf 44}, 1115 (1980).
\bibitem{fadd} L. Faddeev and A. Niemi, Phys. Rev. Lett.
{\bf 82}, 1624 (1999); Phys. Lett. {\bf B449}, 214 (1999).
\bibitem{shab}S. Shabanov, Phys. Lett. {\bf B458}, 322 (1999);
{\bf B463}, 263 (1999); H. Gies, Phys. Rev. {\bf D63}, 125023 (2001).
\bibitem{zucc} R. Zucchini, Int. J. Geom. Meth. Mod. Phys. {\bf 1},
813 (2004).
\bibitem{prd01} W. S. Bae, Y. M. Cho, and S. W. Kim, Phys. Rev. {\bf D65},
025005 (2001).
\bibitem{dewitt}B. de Witt, Phys. Rev. {\bf 162}, 1195 (1967);
1239 (1967).
\bibitem{pesk} See for example, C. Itzikson and J. Zuber, 
{\it Quantum Field Theory} (McGraw-Hill) 1985;
M. Peskin and D. Schroeder, {\it An Introduction to Quantum Field Theory} 
(Addison-Wesley) 1995; S. Weinberg, {\it Quantum Theory of Fields} 
(Cambridge University Press) 1996.
\bibitem{plb82} Y. M. Cho, Phys. Lett. {\bf B115}, 125 (1982).
\bibitem{gold} A. Goldhaber, Phys. Rev. Lett. {\bf 36}, 1122 (1976);
E. Witten, Nucl. Phys. {\bf B223}, 422 (1983).  
\bibitem{tsai} W. Tsai and A. Yildiz, Phys. Rev. {\bf D4}, 3643 (1971); 
T Goldman and W. Tsai, {\it ibid}, 3648 (1971).
\bibitem{table} See, for example, I. Gradshteyn and I. Ryzhik,
{\it Table of Integrals, Series, and Products}, edited by A. Jeffery
(Academic Press) 1994; M. Abramowitz and I. Stegun,
{\it Handbook of Mathematical Functions}, (Dover) 1970.
\bibitem{prd02} Y. M. Cho and D. G. Pak, Phys. Rev. {\bf D65}, 
074027 (2002); Y. M. Cho, H. W. Lee, and D. G. Pak,
Phys. Lett. {\bf B525}, 347 (2002).
\bibitem{jhep05} Y. M. Cho, M. L. Walker, and D. G. Pak, JHEP {\bf 05},
073 (2004); Y. M. Cho and M. L. Walker, Mod. Phys. Lett {\bf A19},
2707 (2004).
\bibitem{schw} J. Schwinger, Phys. Rev. {\bf 82}, 664 (1951).
\bibitem{prl01} Y. M. Cho and D. G. Pak, Phys. Rev. Lett. {\bf 86},
1947 (2001); {\bf 91}, 039151; W. S. Bae, Y. M. Cho, and D. G. Pak, 
Phys. Rev. {\bf D64}, 017303 (2001).
\bibitem{ytmu} Y. M. Cho and D. G. Pak, hep-th/0006051. 
See also Y. M. Cho and D. G. Pak, in
{\it Proceedings of TMU-Yale Symposium on Dynamics
of Gauge Fields}, edited by T. Appelquist and
H. Minakata (Universal Academy Press, Tokyo) (1999).
\bibitem{sch} V. Schanbacher, Phys. Rev. {\bf D26}, 489 (1982).
\bibitem{wil} D. Gross and F. Wilczek, Phys. Rev. Lett. {\bf 26},
1343 (1973); H. Politzer, Phys. Rev. Lett. {\bf 26}, 1346 (1973).
\bibitem{plb06} Y. M. Cho and D. G. Pak, Phys. Lett. {\bf B 525}, 347 (2006).
\bibitem {seib} N. Seiberg and E. Witten, Nucl. Phys. {\bf B426}, 19 (1994);
{\bf B431}, 484 (1994).
\bibitem{cho1} Y. M. Cho, D. G. Pak, and Pengming Zhang, to be published.
\end{thebibliography}
\end{document}